\documentclass[prc,twocolumn,showpacs,preprintnumbers,amsmath,amssymb]{revtex4}

\usepackage{amsmath,amssymb,multirow,epsfig,bm}

\newcommand{\beq}{\begin{equation}}
\newcommand{\eeq}{\end{equation}}
\newcommand{\bea}{\begin{eqnarray}}
\newcommand{\eea}{\end{eqnarray}}

\newcommand{\benn}{\begin{displaymath}}
\newcommand{\eenn}{\end{displaymath}}
\begin{document}

\title{Nuclear tetrahedral configurations at spin zero}

\author{Krzysztof Zberecki$^1$, Paul-Henri Heenen$^2$ and Piotr Magierski$^1$}
\affiliation{$^1$Faculty of Physics, Warsaw University of Technology, ul. Koszykowa 75, 00-662 Warsaw, Poland}
\affiliation{$^2$Service de Physique Nucl\'{e}aire Th\'{e}orique, U.L.B - C.P. 229, B 1050 Brussels, Belgium}

\date{\today}

\begin{abstract}
The possibility of the existence of stable tetrahedral
deformations at spin zero is investigated using the Skyrme-HFBCS
approach and the generator coordinate method (GCM). The study is
limited to nuclei in which the tetrahedral mode has been predicted
to be favored on the basis of non self-consistent models. Our
results indicate that a clear identification of tetrahedral
deformations is unlikely as they are strongly mixed with the axial
octupole mode. However, the excitation energies related to the
tetrahedral mode are systematically lower than those of the axial
octupole mode in all the nuclei included in this study.
\end{abstract}


\maketitle

\section{INTRODUCTION}

Exotic shapes of the nuclear density have always attracted the
interest of physicists. In this respect, the octupole degree of
freedom has played a special role. Axial octupole deformations are
well established, both experimentally and theoretically~\cite{bn},
in several regions of the nuclear chart. It has also been shown
that non axial octupole shapes are competitive with the axial ones
in specific nuclei~\cite{ska92}. However, octupole deformations
are more subtle than quadrupole ones. Stable static octupole
deformations correspond usually to shallow minima as a function of
the octupole deformation~\cite{bfh87,er89,eb03}. Dynamical studies
have shown that octupole correlations in the ground state manifest
themselves predominantly by a spreading of the wave function
around the left-right symmetric mean-field
configuration~\cite{mbw95}.

Nuclear tetrahedral deformations have been recently investigated
in Refs.\cite{ld,dgs,dgs2}. Symmetry arguments show that
tetrahedral deformations induce a 4-fold degeneracy of the
single-particle spectra. Based on this peculiar shell structure,
the stability of such configurations has been conjectured in
specific nuclei. Such a large degeneracy results indeed in large
gaps in the single-particle spectrum  and increases the shell
effects for specific values of the neutron and proton numbers.
Those values have been dubbed 'tetrahedral magic numbers' and have
been determined using a microscopic-macroscopic model based on a
Woods-Saxon potential\cite{ld,dgs,dgs2,sod,dcd,ddd}.

However these conjectures almost exclusively originate from
approaches based on non self-consistent average nuclear
potentials, with only a limited support of self-consistent
calculations. It is therefore clearly necessary to test their
validity in the framework of up-to-date theories using modern
energy density functionals. We  address this issue in the present
paper in the framework of self-consistent mean-field methods using
Skyrme interactions. Since to go beyond a pure mean-field approach
has been shown to be crucial for octupole deformations, we have
also studied the stability of tetrahedral shapes using the
generator coordinate method (GCM). This study is focused on nuclei
in which the tetrahedral mode has been predicted to be favored:
$^{80,98,110}$Zr, $^{152-156}$Gd and $^{160}$Yb. It extends our
previous study which was limited to$^{80,98}$Zr\cite{zber1,zber2}
(Ref. I) by considering the dynamical coupling between the axial octupole
and the tetrahedral degrees of freedom. Since tetrahedral
shapes are generated by the non-axial intrinsic octupole moment
$Q_{32}\propto r^{3}( Y_{3 2} + Y_{3 -2})$ it is likely that they
are in strong competition with axial octupole shapes. Most details
about our method can be found in Ref. I, those on the GCM in
Ref.~\cite{gcm} and on its application to 2-dimensional octupole
calculations in Ref.~\cite{shb93a}.

%\emph{Changes}

 Our aim in this study is to determine whether there
are situations in which a configuration can be unambiguously
identified as tetrahedral. We will therefore first identify which
are the possible coexisting structures in the nuclei for which
tetrahedral deformations have been predicted. We will then study
whether some of these configurations provide clear signatures of
tetrahedral shapes.

\section{MEAN-FIELD CALCULATIONS AND PARITY PROJECTION}

Octupole deformations of the nuclear density are generated by
introducing in the HFBCS equations  the axial $Q_{30}\propto
r^{3}Y_{3 0}$ and the triaxial  $Q_{32}\propto r^{3}( Y_{3 2} +
Y_{3 -2})$ moments as constraining operators. This last one is the
operator generating tetrahedral deformations. The details of the
HFBCS calculations have been described in Ref. I. The pairing
interaction strength has been adjusted to reproduce 'experimental
pairing gaps' in the same way as described in Ref. I. In particular
the standard prescription based on the odd-even difference of binding
energies has been used~\cite{dmn}.
The pairing strength for $^{98}$Zr and $^{110}$Zr
has been adjusted to reproduce the pairing gap in $^{102}$Zr.
In Gd isotopes the pairing strength has been adjusted to reproduce
pairing gaps in $^{154}$Gd. For studies of $^{80}$Zr and $^{160}$Yb
the pairing strength has been adjusted for each of these nuclei individually.
These reference nuclei have been chosen
in order to minimize the influence of deformation changes for
the determination of the pairing gap.

%\emph{Changes}

The experimental data suggest that all the nuclei in which
tetrahedral deformations have been predicted theoretically have a
well deformed ground state (with the possible exception of
$^{98}$Zr). A first question to address is whether a spherical
configuration may coexist at low excitation energy in any of the
studied nuclei. As we shall see in the following, the single-particle 
pattern typical of a tetrahedral
configuration is very well preserved close to sphericity. 
Therefore a 'tetrahedral magic' nucleus in which
the spherical configuration is at low energy represents
a particularly promising case 
of the existence of tetrahedral configurations. Let us
first focus on the case of $^{110}$Zr for which the variation of
the energy as a function of the axial quadrupole deformation is
shown in Fig.~1. Octupole deformations are set to zero for all
values of the quadrupole moment. The calculations have been
performed for three representative Skyrme parametrizations. A
deformed prolate ground state at large deformation is obtained for
all three interactions. It coexists with an oblate minimum and, in
the case of SLy4 and SkM$^{*}$, with a local minimum for the
spherical configuration. It is only for the SLy4 parametrization
that the spherical and the prolate minima are almost degenerate.

The case of $^{110}$Zr illustrates well the energy dependence on
the quadrupole degree of freedom obtained for all studied nuclei
with the three Skyrme forces. The ground state corresponds in all
cases to a deformed prolate minimum (with the only exception of
$^{80}$Zr for the SLy4 parametrization but this is in clear
contradiction with the experimental data~\cite{lcc87}). The
spherical configurations are excited by a few MeV, except in some
Zr isotopes where depending on the Skyrme parametrization they are
obtained at excitation energies smaller than 1~MeV.

\begin{figure}[h]\label{fig_zr100}
\vspace*{-0.3cm}
\includegraphics[scale=0.3,angle=-90]{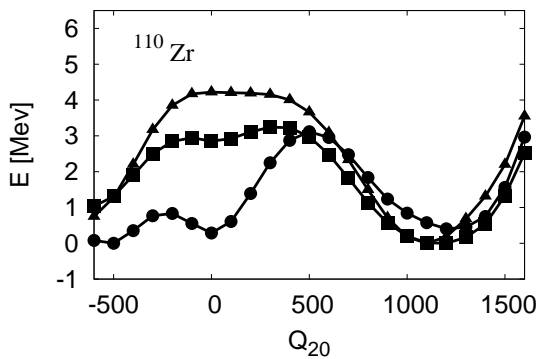}
\vspace*{0cm}
\caption{\label{fig1}
Variation of the total mean-field energy as a function of the
quadrupole $Q_{20}$ moment (in $fm^2$). Triangles, squares and
circles denote results obtained using the SIII, SkM$^{*}$ and SLy4
parameterizations of the Skyrme force, respectively }
\vspace*{-0.4cm}
\end{figure}

A first question that must be answered is how this picture is
affected by tetrahedral deformations and whether the magnitude of
the energy gain obtained  around the spherical minimum is large
enough to bring its energy close to that of the deformed ground
state.

%\emph{Changes ended}

 The variation of the mean-field energy with
octupole deformations when the quadrupole moment is constrained to
be zero is shown in Fig.~2 for the six nuclei that we have
selected. These results have been obtained with the SLy4 Skyrme
parametrization and their dependence on the choice of the
parametrization is illustrated by results obtained for $^{110}$Zr
with SIII. Note that, contrary to Fig.~1, the HFBCS results are
very similar for both interactions. We have also checked for
several nuclei and for SkM$^{*}$ that the behaviour of the HFBCS
energy as a function of the octupole degrees of freedom is
qualitatively similar in all cases.

\begin{figure}[H]
  \begin{center}
    \begin{tabular}{cc}
      \resizebox{70mm}{!}{\includegraphics[angle=270]{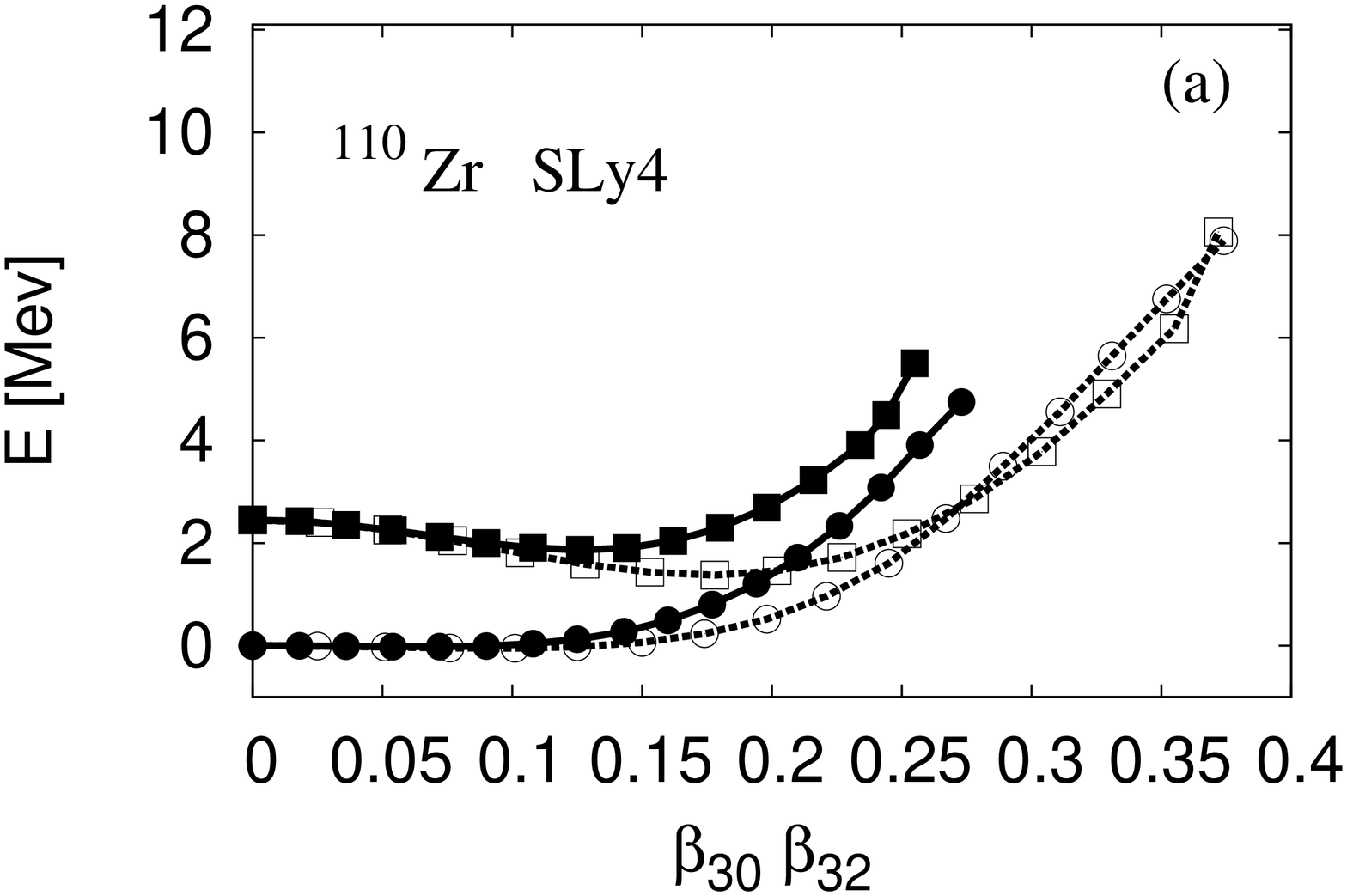}} &
      \resizebox{70mm}{!}{\includegraphics[angle=270]{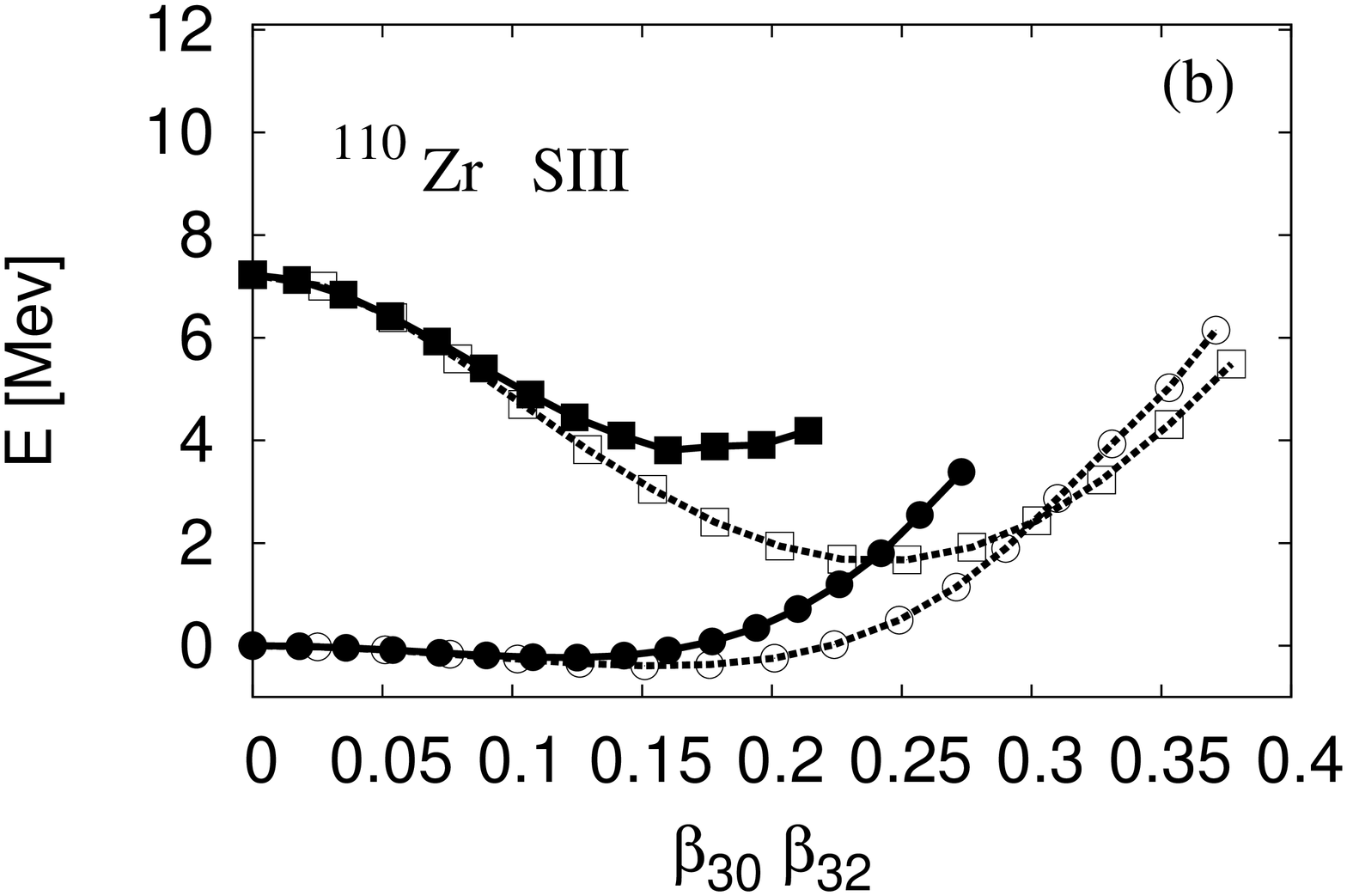}} \\
      \resizebox{70mm}{!}{\includegraphics[angle=270]{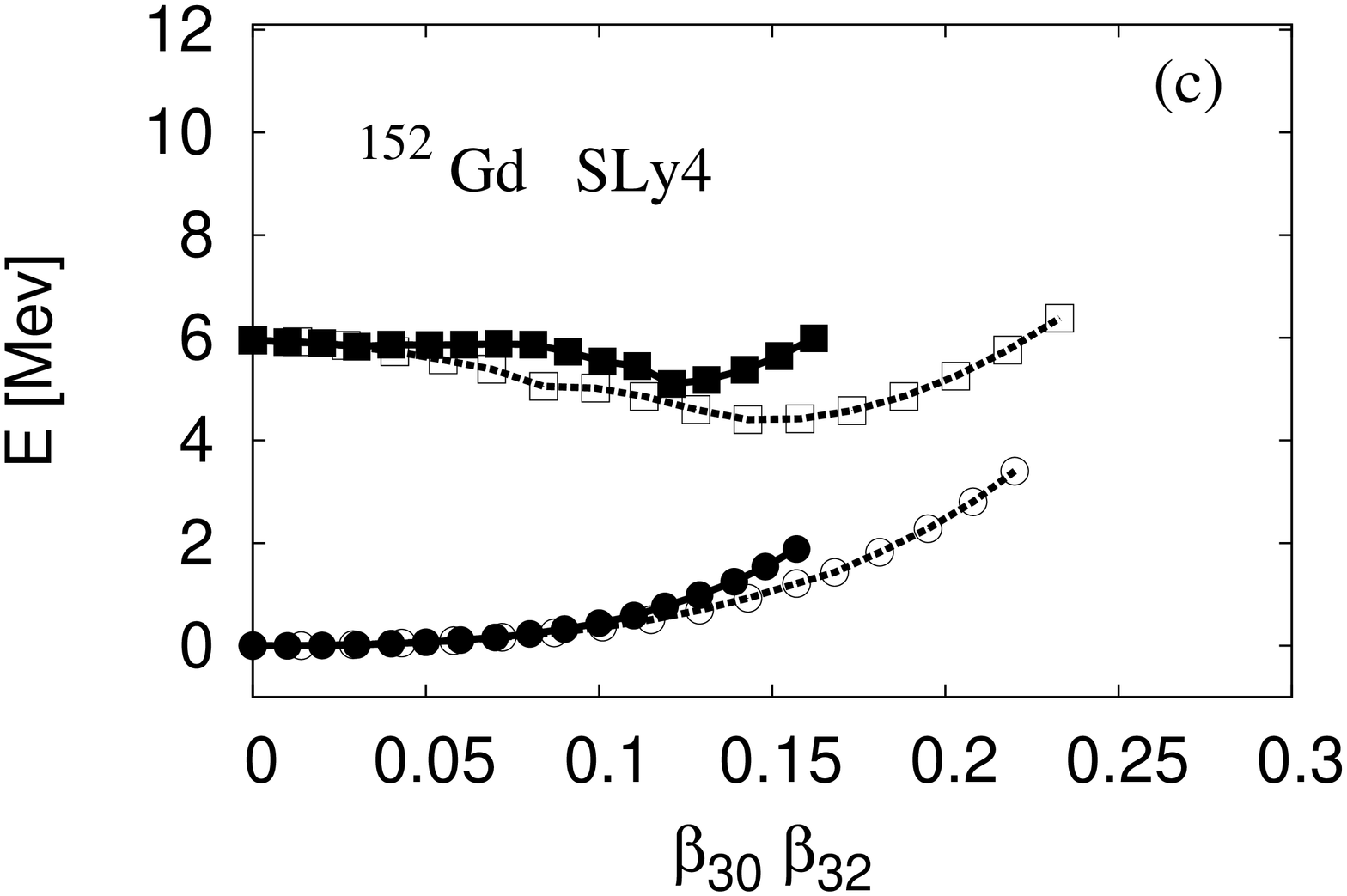}} &
      \resizebox{70mm}{!}{\includegraphics[angle=270]{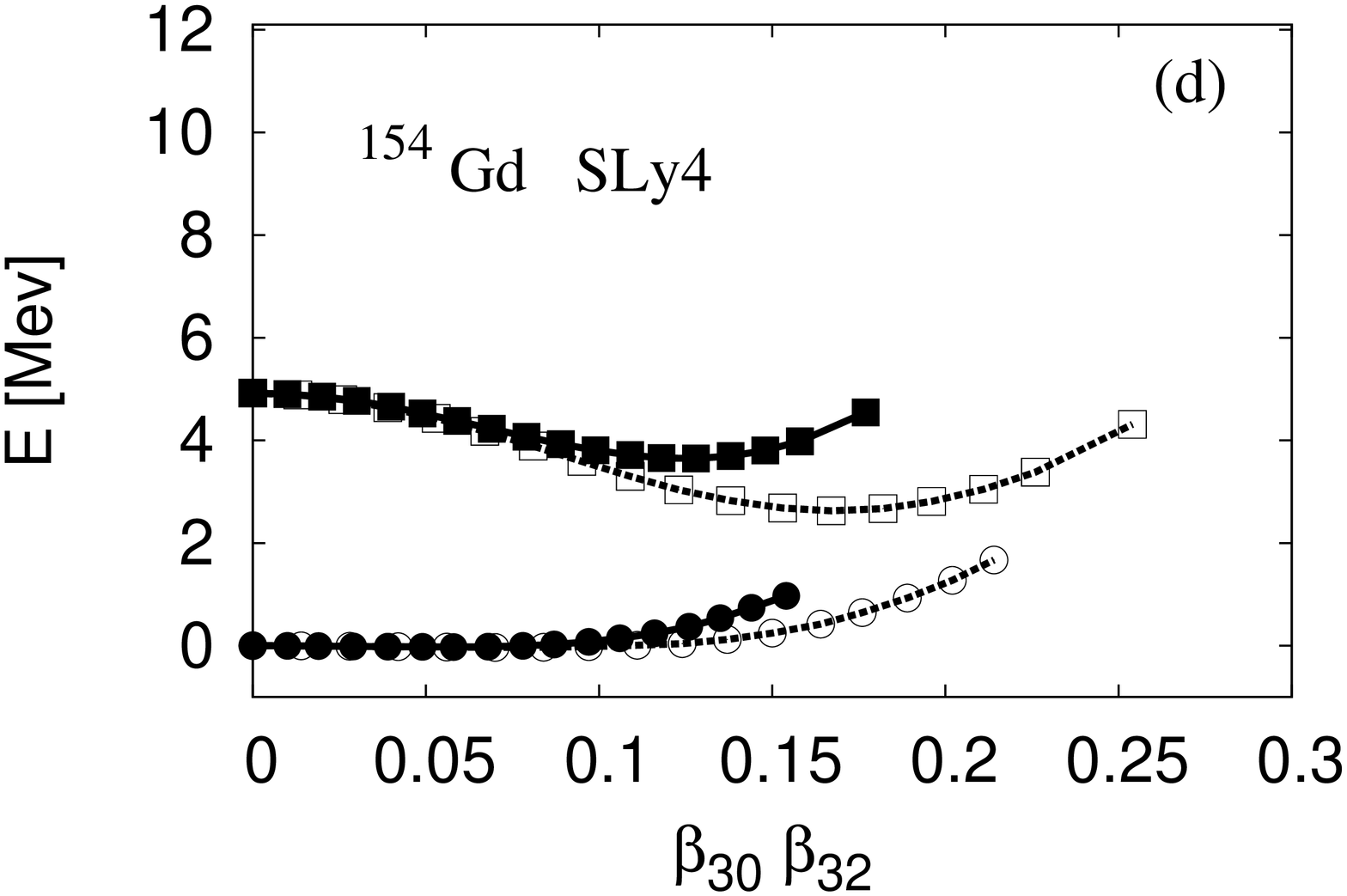}} \\
      \resizebox{70mm}{!}{\includegraphics[angle=270]{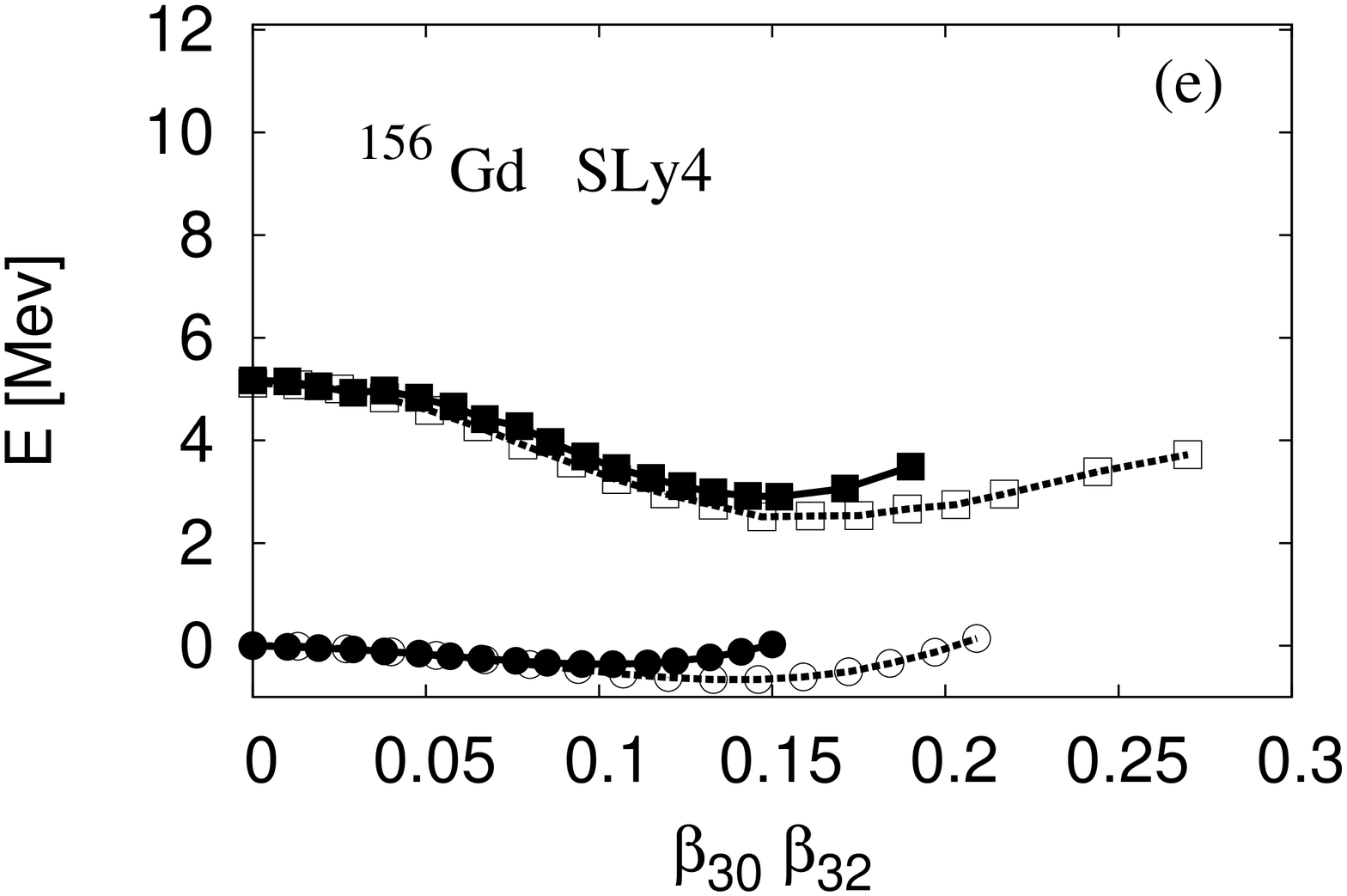}} &
      \resizebox{70mm}{!}{\includegraphics[angle=270]{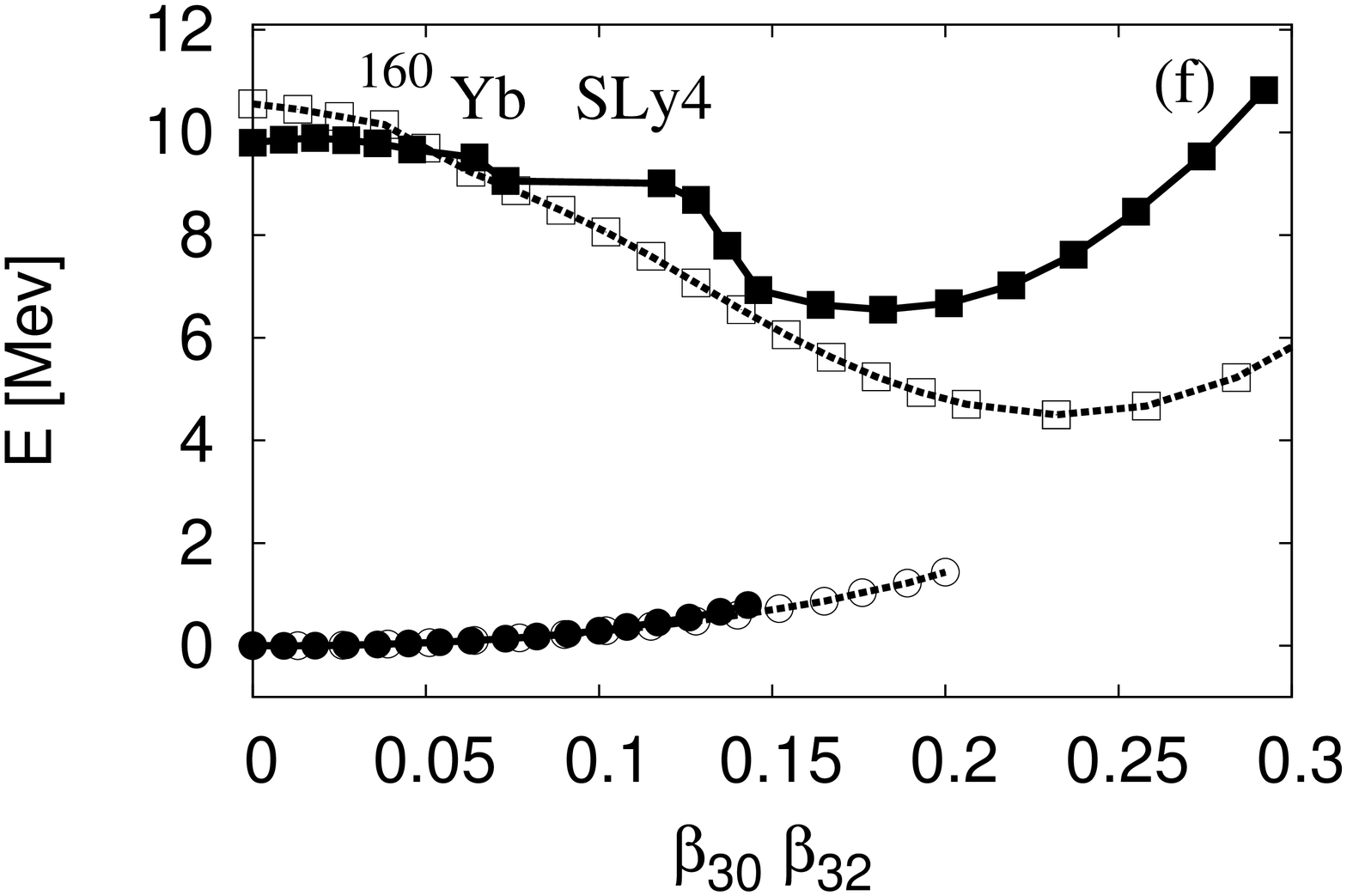}} \\
    \end{tabular}
    \caption{ Total mean-field energy obtained with the HF (squares) and HFBCS (circles) methods
              as a function of the dimensionless $\beta_{30}$ (filled symbols)
              and $\beta_{32}$ (open symbols) deformation parameters.}
    \label{hfbcs3}
  \end{center}
\end{figure}

\begin{figure}[H]
  \begin{center}
    \begin{tabular}{cc}
      \resizebox{70mm}{!}{\includegraphics[angle=270]{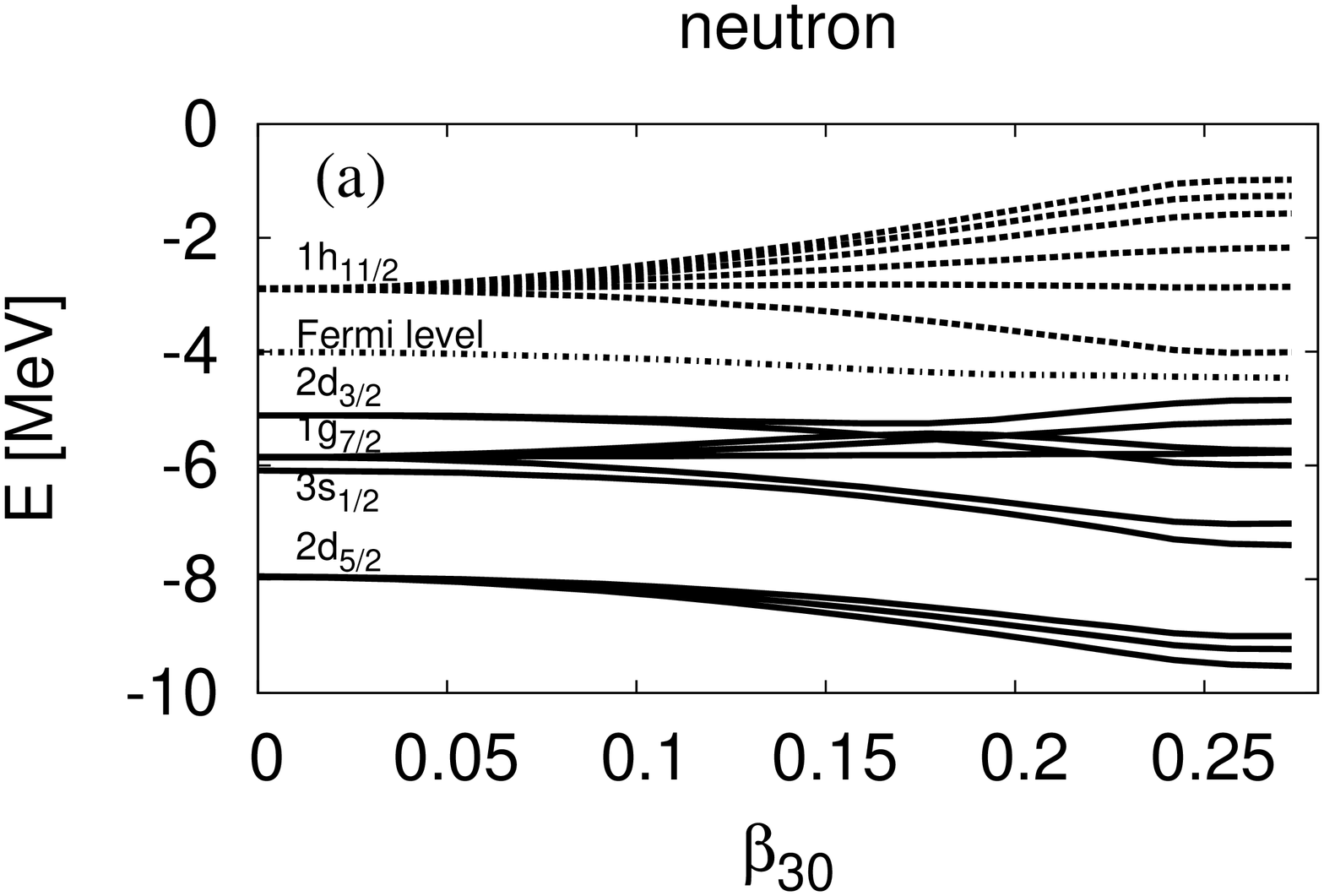}} &
      \resizebox{70mm}{!}{\includegraphics[angle=270]{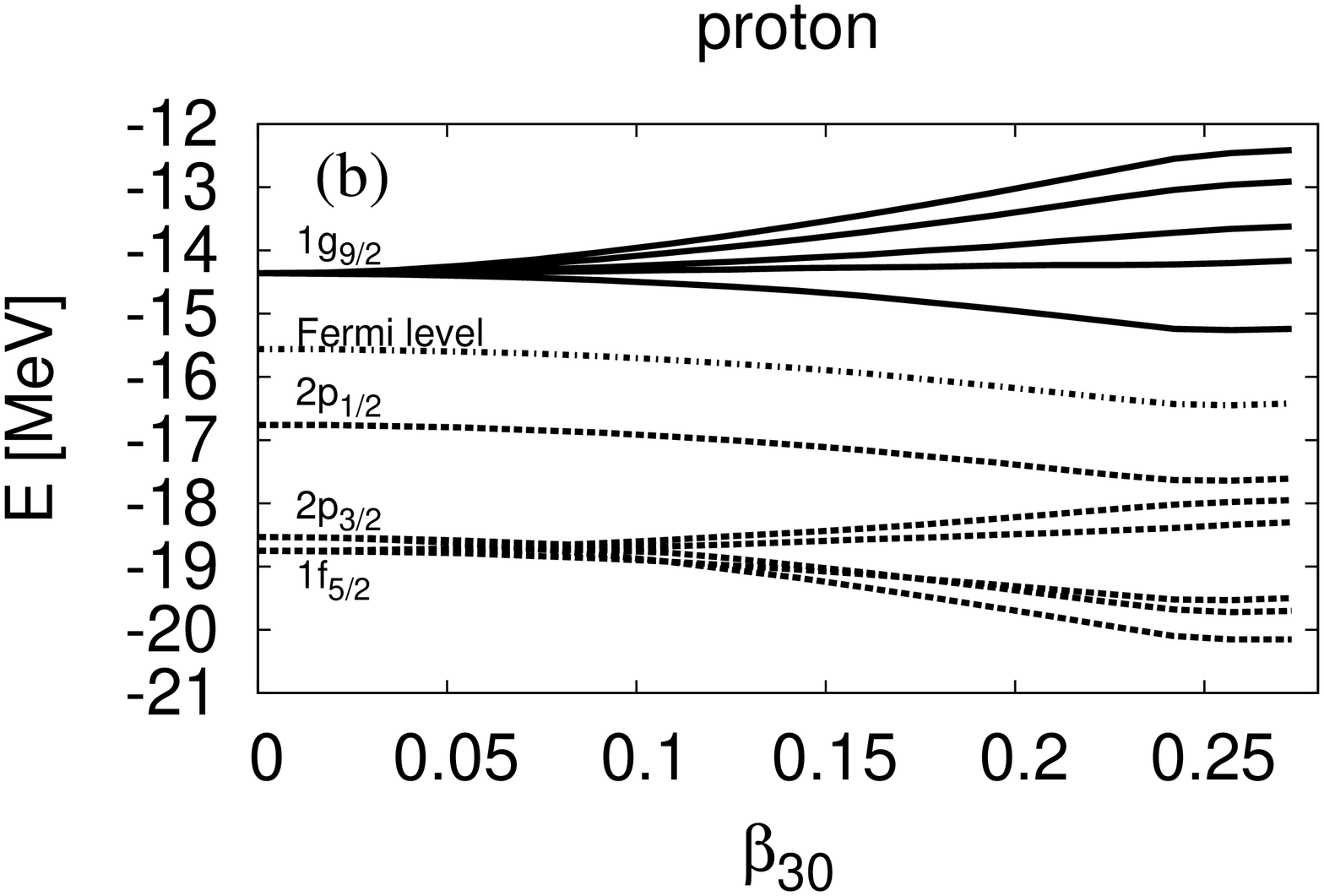}} \\
      \resizebox{70mm}{!}{\includegraphics[angle=270]{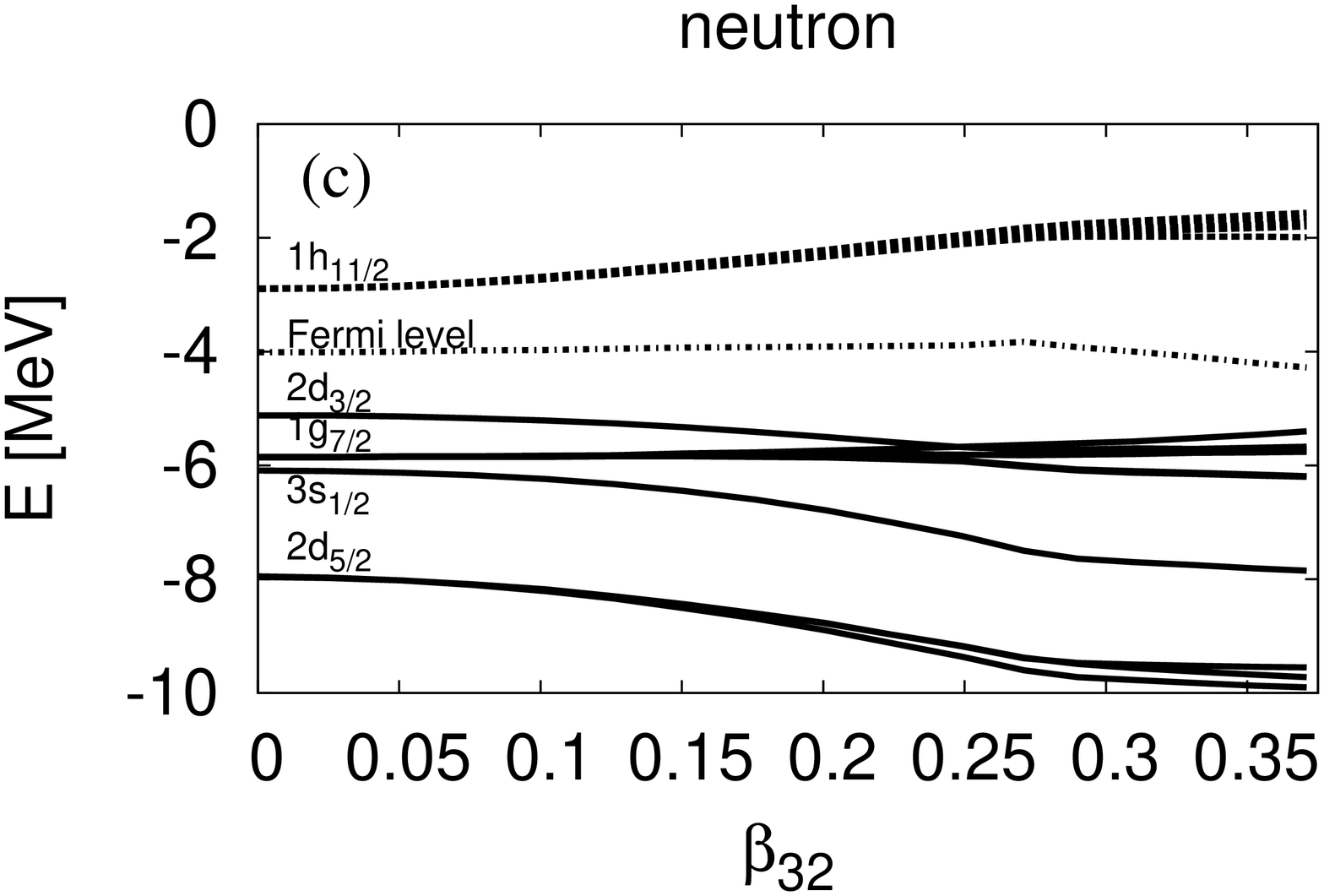}} &
      \resizebox{70mm}{!}{\includegraphics[angle=270]{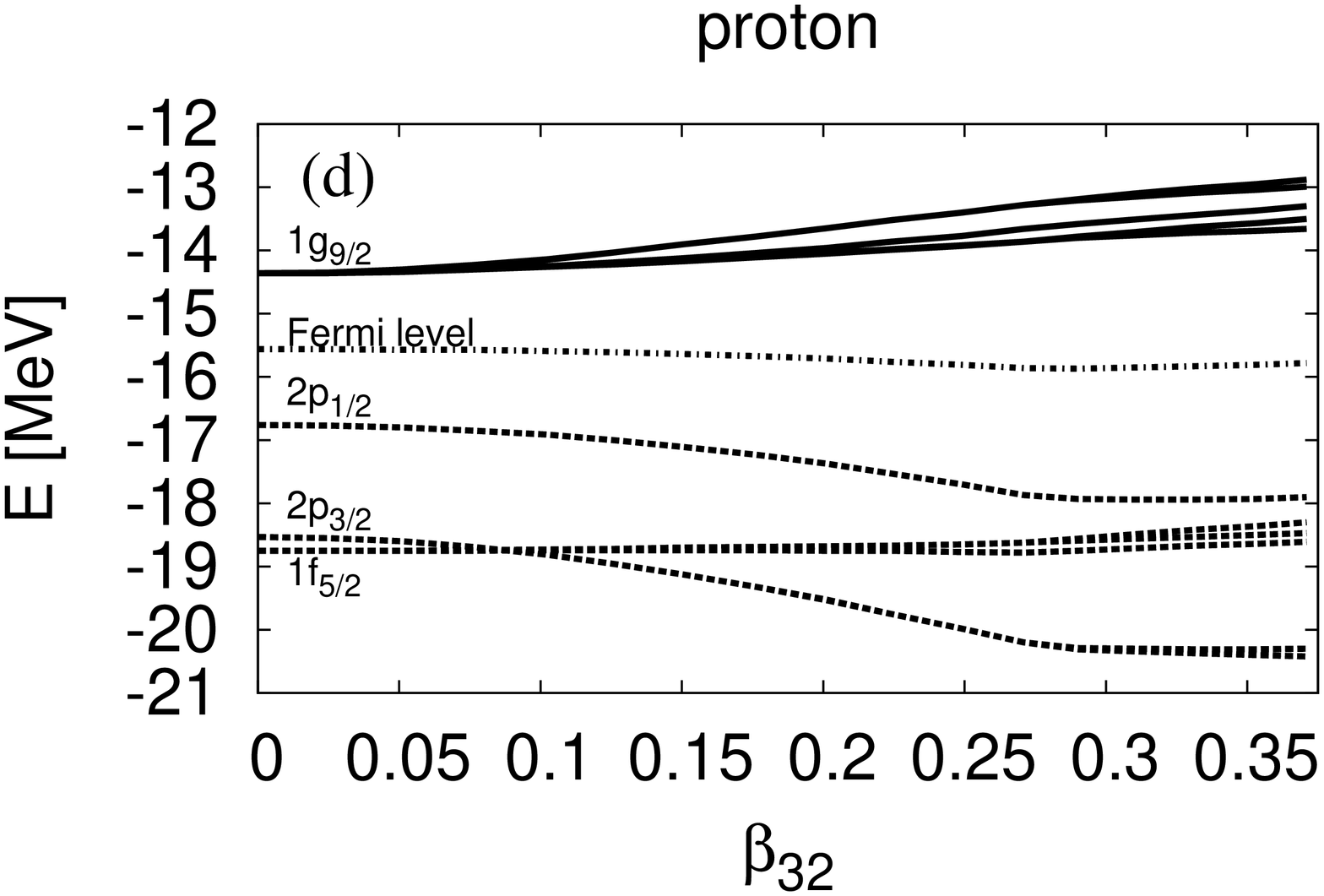}} \\
    \end{tabular}
    \caption{ Single-particle energies as a function of octupole deformation $\beta_{30}$
     and $\beta_{32}$ for $^{110}$Zr calculated for SLy4 force. The positive and negative parity
     levels are denoted by solid and dashed lines, respectively.}
    \label{sp_levels}
  \end{center}
\end{figure}
The octupole deformations are parameterized by dimensionless
parameters proportional to the octupole moments (see Ref. I). There is no
direct connection between
the values of $\beta_{30}$  and $\beta_{32}$ and one
must be cautious when comparing axial and tetrahedral
deformation energy curves. The only nucleus for
which a sizeable tetrahedral minimum is obtained is $^{156}$Gd with
an energy gain of 500~keV with respect to the spherical
configuration. The energy is rather flat as a function of both
$\beta_{30}$  and $\beta_{32}$ for the other nuclei, although the
meaning of this shallowness can only be made precise by dynamical
calculations. The energy curve obtained for $^{110}$Zr with SIII
is only marginally different from that of SLy4, with a tetrahedral
minimum 100~keV below the spherical point.

These mean-field calculations have been performed using a single
constraint on either the $Q_{30}$ or $Q_{32}$ moments. Since the
unconstrained degrees of freedom are completely relaxed (except
for the quadrupole moment and either $Q_{30}$ or $Q_{32}$ which
were set to zero) by the
variational procedure, there is no guarantee that the
unconstrained moments remain equal to zero (except those
that are forbidden due to the imposed symmetry conditions
 - see Ref. I) for all values of the constraint. Nevertheless, the behavior typical of the tetrahedral
symmetry is largely preserved up to a deformation of
$\beta_{32}\approx 0.2-0.3$. Up to these values, the
single-particle states exhibits the 4-fold degeneracies
characteristic for the point group $T_{d}^{D}$. In general, the
single particle energies as a function of $Q_{32}$ exhibit more
bunching as compared to the $Q_{30}$ direction. However, this does
not translate into a lower energy for the tetrahedral
configuration.

As a typical example, the variation of the single-particle
energies as a function of octupole deformations is shown in Fig.~3
for $^{110}$Zr. One can see that the single particle states are
still almost degenerate at $\beta_{32}$ equal to 0.15 but not at
$\beta_{30}$ equal to 0.15, which is in both cases the deformation
beyond which the mean-field energy starts to increase. Note also
that both the tetrahedral and the spherical configurations have a
similar single-particle structure since there are no level crossings
between these configurations. The same is true for the axial
octupole configuration. Moreover the HFBCS calculations indicate
almost no barrier between tetrahedral and axial octupole minima.
These facts indicate that all the three configurations may be
strongly mixed when the octupole collective dynamics is taken into
account (see next section).

Since tetrahedral deformations break parity, projection on parity
gives rise to an energy gain for the positive parity as soon as
the octupole moments have a non zero value and it generates a
distinct energy curve for the negative parity. We have restored
both particle number and parity by projecting the mean-field wave
functions. The projected potential energy is defined as:
\beq
E(N,Z,\beta_{3\mu})_{\pm} = \frac{\langle
\phi(\beta_{3\mu})|\hat{H}\hat{P}_{(\pm, N,
Z)}|\phi(\beta_{3\mu})\rangle} {\langle
\phi(\beta_{3\mu})|\hat{P}_{(\pm, N,
Z)}|\phi(\beta_{3\mu})\rangle} ,
\eeq
where $|\phi(\beta_{3\mu})\rangle$ are HFBCS wave functions
generated with the constraint $\langle
\phi(\beta_{3\mu})|\hat{Q}_{3\mu}|\phi(\beta_{3\mu})\rangle=C_{\mu}\beta_{3\mu}$,
where $C_{0}=\displaystyle{\frac{3}{4 \pi}} A^{2} r_{0}^{3},
C_{2}=C_{0}/\sqrt{2}$ with $A=N+Z$ and $r_{0}=1.2 fm$ (see Ref. I).
The operator $\hat{P}_{(\pm,N,Z)}$ is the product of operators
projecting  on $\pi = \pm 1$ parity  and on $N$ neutrons and $Z$
protons. The parity-projected energies are shown in Fig.~4, except
for $^{80}$Zr and $^{98}$Zr which were already discussed
in Ref. I.

\begin{figure}[H]\label{fig4}
  \begin{center}
    \begin{tabular}{cc}
      \resizebox{70mm}{!}{\includegraphics[angle=270]{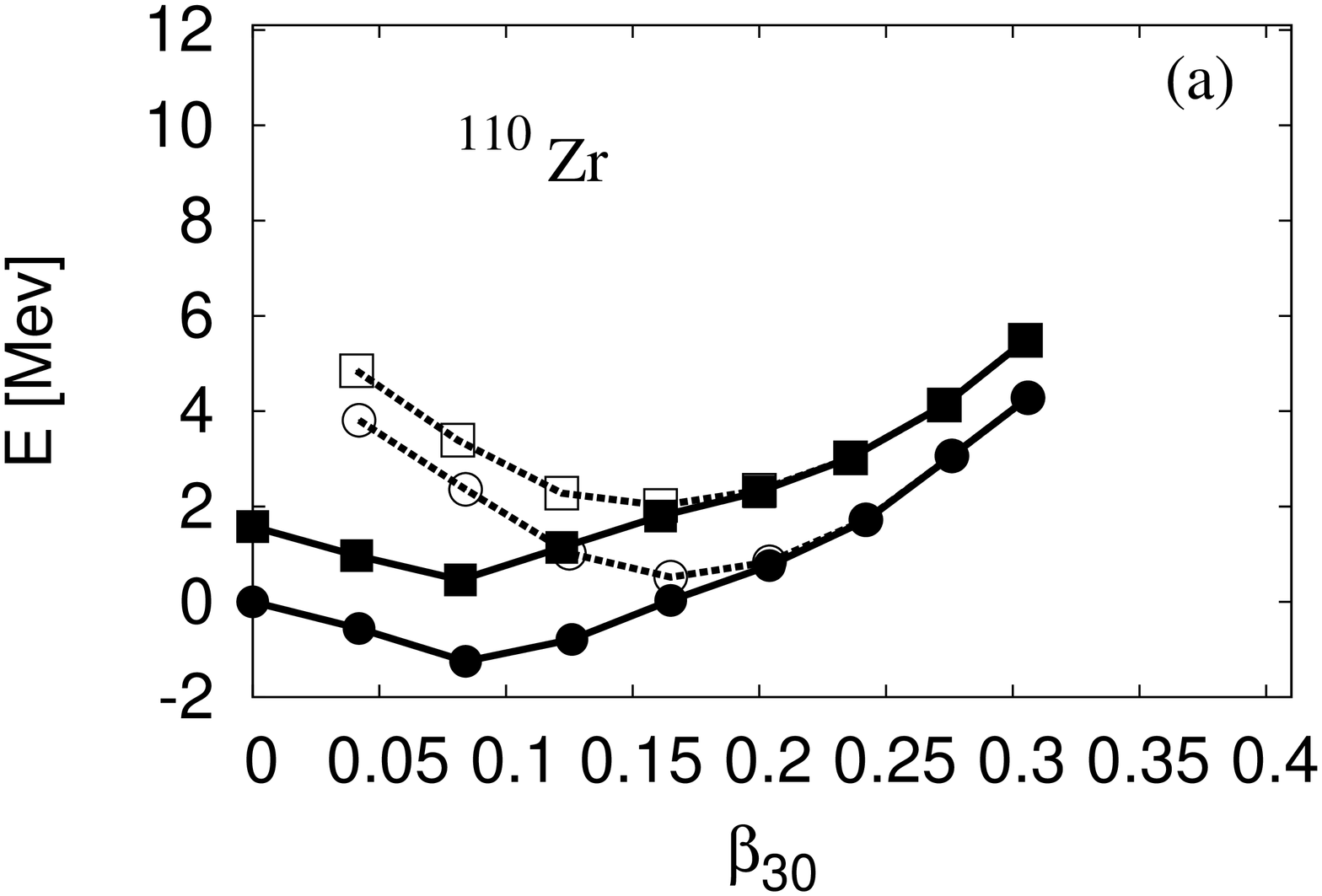}} &
      \resizebox{70mm}{!}{\includegraphics[angle=270]{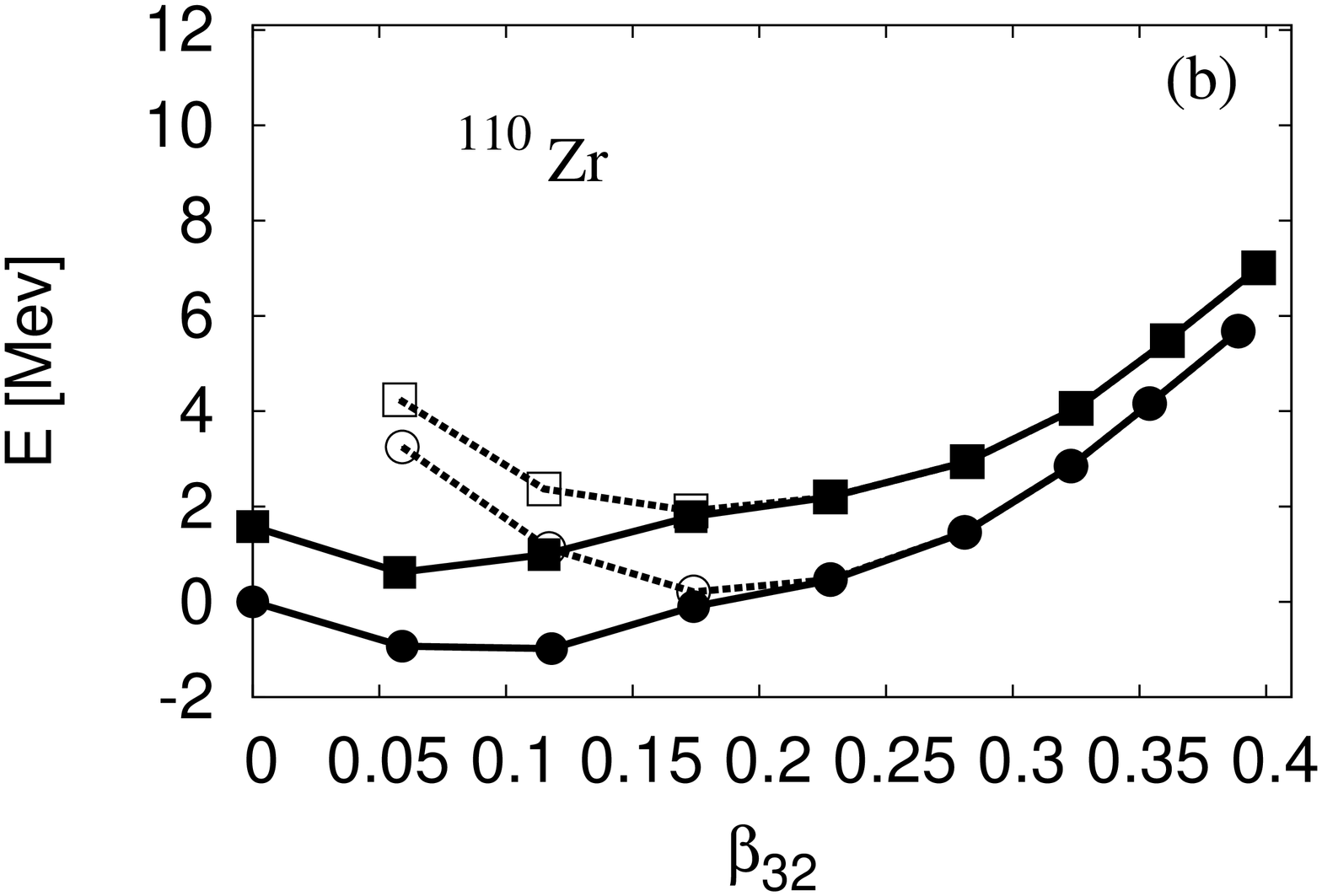}} \\
      \resizebox{70mm}{!}{\includegraphics[angle=270]{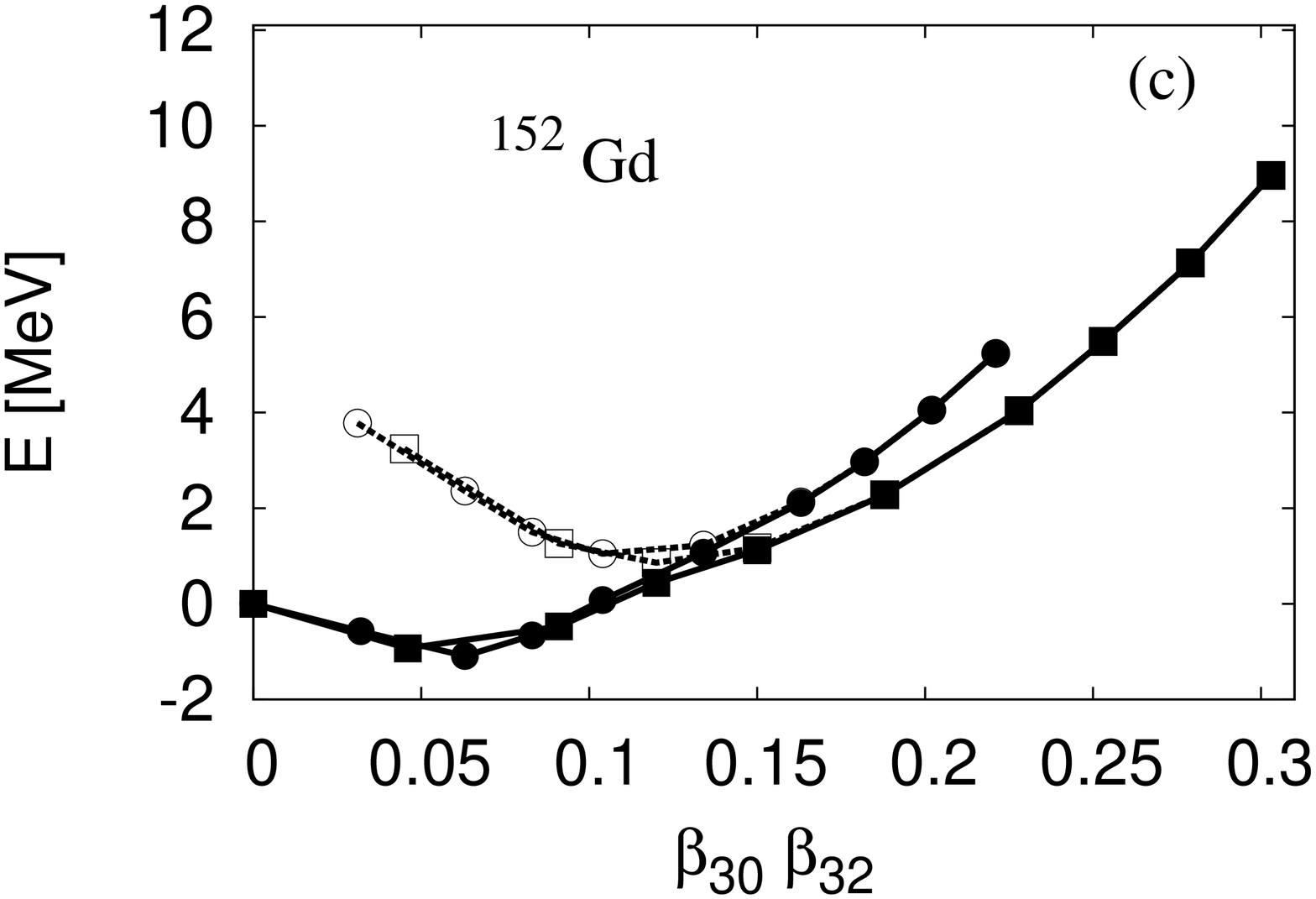}} &
      \resizebox{70mm}{!}{\includegraphics[angle=270]{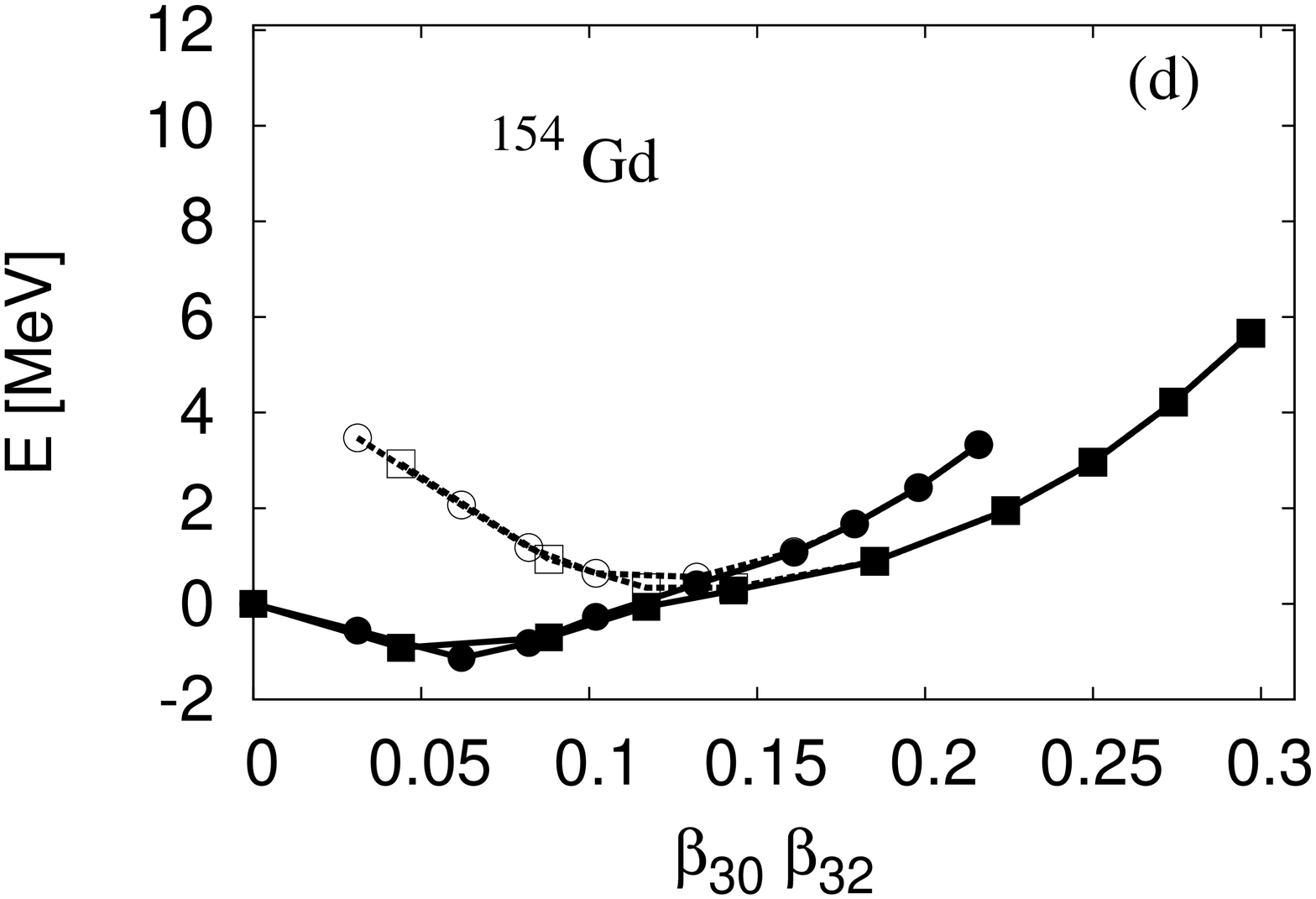}} \\
      \resizebox{70mm}{!}{\includegraphics[angle=270]{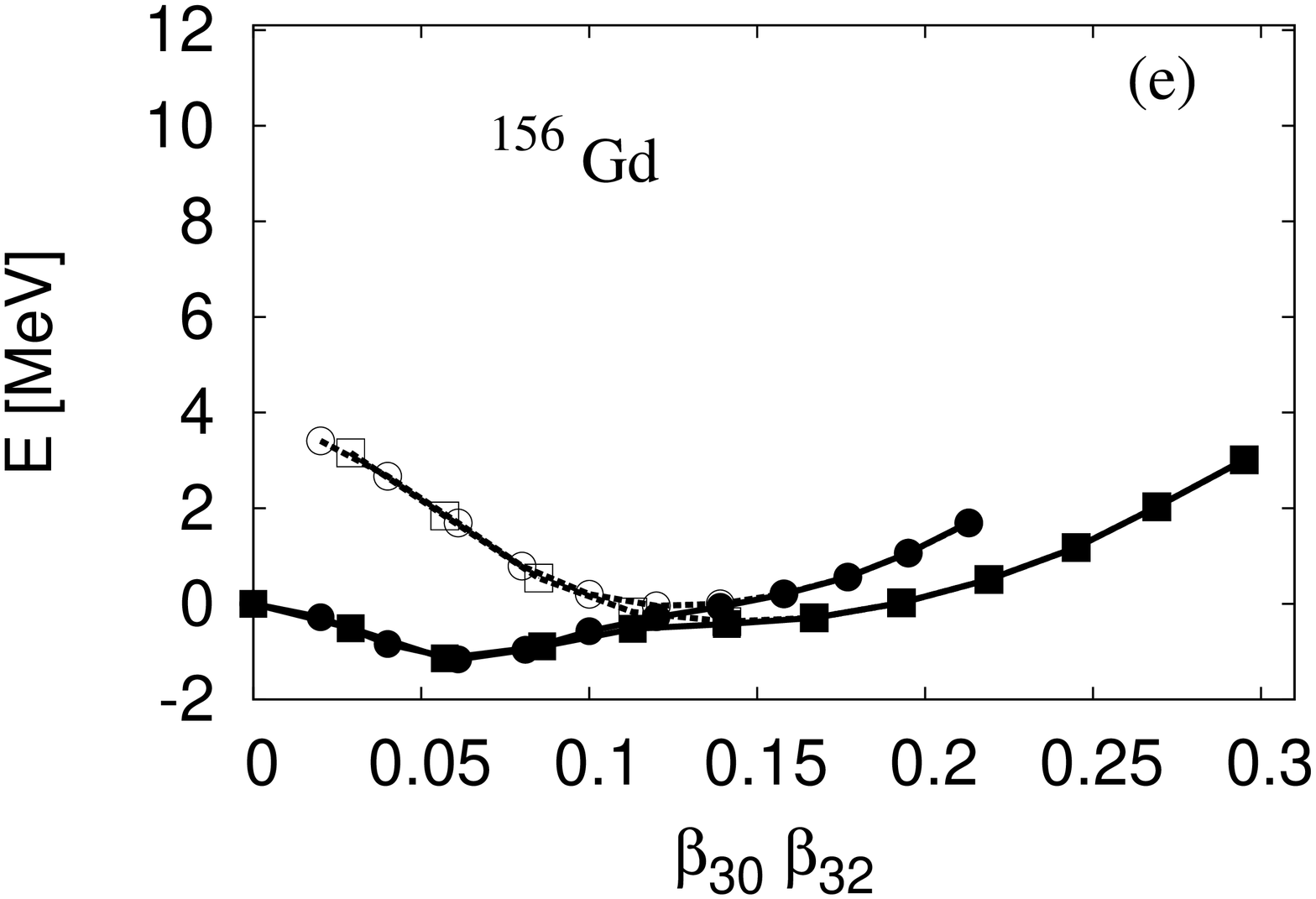}} &
      \resizebox{70mm}{!}{\includegraphics[angle=270]{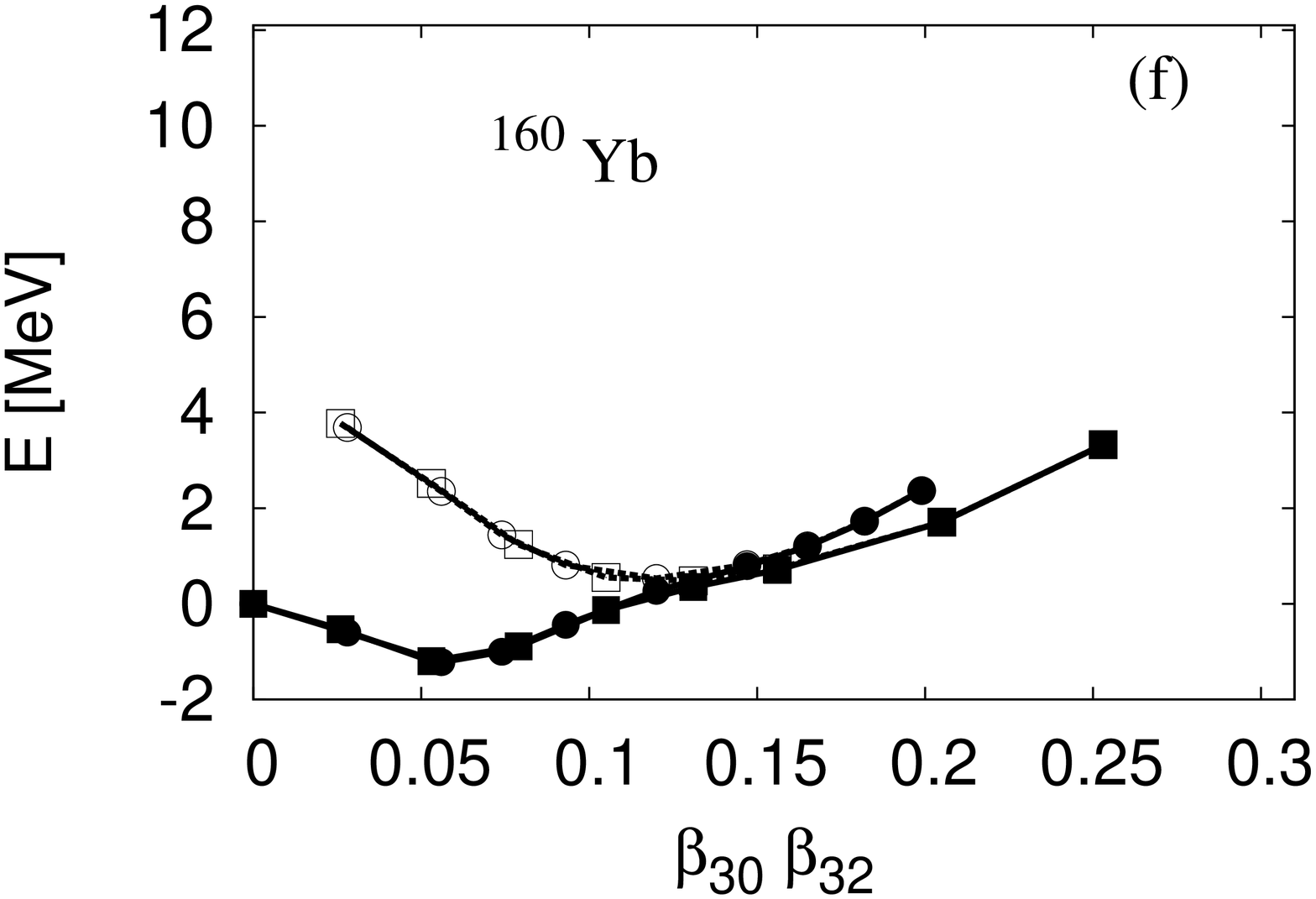}} \\
    \end{tabular}
    \caption{ Parity and particle number projected energies as a function of the octupole
    deformations $\beta_{30}$ and $\beta_{32}$. In the two top subfigures (a and b)
corresponding to $^{110}$Zr,
the results obtained with the SLy4 (circles) and SIII  (squares) are compared. In
the other subfigures, only calculations performed with SLy4 are
shown. The circles and squares denote then the energies as a
function of $\beta_{30}$ and $\beta_{32}$, respectively. In all subfigures filled
and open symbols refer to positive and negative parity, respectively.}
    \label{ep_1d_2}
  \end{center}
\end{figure}

The situation is similar to that discussed already in
Ref.\cite{zber1,zber2}. The positive parity curves exhibit a small
minimum for nonzero values of $\beta_{30}$ and $\beta_{32}$. The
energy gain due to the parity restoration is of the order of 1 MeV
for both axial octupole and tetrahedral deformations (see Fig.4).
In fact for all nuclei considered (with the exception of
$^{160}$Yb ) the axial octupole minimum has a slightly lower
energy than the tetrahedral one (see table I).
\begin{table}[H]
\begin{center}
\begin{tabular}{ccccccc}
\hline
\multicolumn{1}{c}{} & \multicolumn{2}{c}{SLy4}  & \multicolumn{2}{c}{SIII} \\
\hline
\hline
Sly4 & $\beta_{30}$ & $\beta_{32}$ & $\beta_{30}$ & $\beta_{32}$ \\
\hline
$^{110}$Zr  & -1.25  & -0.98  & -1.12  & -0.96 \\
$^{152}$Gd  & -1.09  & -0.93  &    -   &   -   \\
$^{154}$Gd  & -1.13  & -0.92  &    -   &   -   \\
$^{156}$Gd  & -1.16  & -1.14  &    -   &   -   \\
$^{160}$Yb  & -1.19  & -1.20  &    -   &   -   \\
\hline
\end{tabular}
\end{center}
\caption{Energies (in MeV) of the positive parity configurations with respect to the spherical
configurations for axial ($\beta_{30}$) and tetrahedral ($\beta_{32}$) configurations.}
\label{hw_tab0}
\end{table}

The dependence of energy on octupole deformations is not
significantly modified by the projection on positive parity. An
interesting property of the particle projection is that it makes
the results rather weakly dependent on the pairing strength, in
contrast to pure mean-field calculations. We have checked that the
differences between the energy of the octupole minima and that of
the spherical configuration are not significantly modified by a
variation of the pairing strength within the interval between half
and twice the physical value. This holds for both the positive and
negative parity states. The main effect of an increase of the
pairing strength is an increase of the energy difference between
the positive and negative parity minima.

\section{TWO-DIMENSIONAL GCM}

To be unambiguously identified experimentally, tetrahedral
deformations should have a clear signature, which allows to
distinguish them from axial octupole deformations. The GCM enables
to study the coupling between both octupole modes and to see
whether tetrahedral shapes can be separated from axial octupole
shapes. We have therefore performed two-dimensional GCM
calculations in which the axial and non axial octupole shapes are
coupled. This coupling was not considered in Ref. I, where
separate dynamical calculations were performed along the
collective paths determined by non zero $Q_{30}$ and $Q_{32}$
values, respectively. The method we apply is similar to that
introduced by Skalski et al.~\cite{shb93a}.

%\emph{Changes (move text}

A collective wave function is constructed by mixing the mean-field
states corresponding to different values of the octupole
moments, after their projection on particle numbers:
\beq
|\Psi\rangle = \int f(\beta_{30},\beta_{32}) \hat{P}_{(N,
Z)}|\phi(\beta_{30}, \beta_{32})\rangle d \beta_{30}d \beta_{32}
\eeq
The coefficients $f(\beta_{30},\beta_{32})$ are determined by minimizing
the total energy of the collective wave function $|\Psi\rangle$.

Our collective space forms a plane specified by $Q_{30}$ and  $
Q_{32}$, or equivalently by $\beta_{30},\beta_{32}$. This requires
to consider HFBCS states inside a rectangle specified by
"corners": $(\pm \beta_{30max},\pm \beta_{32max})$. However the
full problem can be decomposed in four subspaces by introducing
combinations of states in the four quadrants. Starting from
$|\phi_1 \rangle = |\phi(\beta_{30},\beta_{32}) \rangle$, one
constructs the four states:
\begin{eqnarray*}
  |\phi_2 \rangle&=& |\phi(-\beta_{30},-\beta_{32})\rangle =\hat{P} |\phi_1\rangle, \\
  |\phi_3\rangle &=& |\phi(-\beta_{30},+\beta_{32})\rangle =\hat{P}_{xy}\hat{P} |\phi_1\rangle, \\
  |\phi_4 \rangle&=& |\phi(+\beta_{30},-\beta_{32})\rangle =\hat{P}_{xy}|\phi_1\rangle,
\end{eqnarray*}
where $\hat{P}$ is the parity operator and $\hat{P}_{xy}$ is the
reflection operation in which $x$ and $y$ coordinates are
exchanged. Thus, one needs to generate only the HFBCS basis in
$1/4$ of the rectangle and extends it to the full square thanks to
these relations. Another interest of this decomposition is that
$\hat{P}$ and $\hat{P}_{xy}$ commute with the Hamiltonian. This
means that they can be used to label GCM  eigenstates. Both
$\hat{P}$ and $\hat{P}_{xy}$ are projectors, so the quantum
numbers associated to each operator take the values $\pm 1$. From
the wave functions $|\phi_{i}\rangle$, $i=1,2,3,4$, one can define
a new basis in which both $\hat{P}$ and $\hat{P}_{xy}$ are
diagonal, namely:
\begin{eqnarray*}
|\Phi_{++} \rangle&=& \frac{1}{2}(|\phi_1\rangle +|\phi_2\rangle +|\phi_3\rangle +|\phi_4\rangle), \\
|\Phi_{--} \rangle&=& \frac{1}{2}(|\phi_1\rangle -|\phi_2\rangle +|\phi_3\rangle -|\phi_4\rangle), \\
|\Phi_{-+} \rangle&=& \frac{1}{2}(|\phi_1\rangle -|\phi_2\rangle -|\phi_3\rangle +|\phi_4\rangle), \\
|\Phi_{+-} \rangle&=& \frac{1}{2}(|\phi_1\rangle +|\phi_2\rangle-|\phi_3\rangle -|\phi_4\rangle),
\end{eqnarray*}
where the first index of $|\Phi_{kl} \rangle$ denotes the
eigenvalue with respect to parity and the second index with
respect to $x-y$ reflection. One can easily check that $|\Phi_{+-}
\rangle$ is identically zero in the absence of either axial or
tetrahedral deformations, while $|\Phi_{-+} \rangle$ and
$|\Phi_{--} \rangle$ are zero when $\beta_{30}$ or  $\beta_{32}$
are zero, respectively. For this
reason, we have dubbed the excited states corresponding to $k=l=-1$
tetrahedral excitations, those corresponding to $k=-1, l=+1$ axial
excitations and those corresponding to $k=1, l=-1$ mixed octupole
excitations.

In this basis $|\Phi_{kl} \rangle$, the Hamiltonian does not
couple states corresponding to different values of $k$ and $l$ and
the GCM equation decomposes into four equations for each set
$(k,l)$. The resulting GCM wave functions are expressed by:
\beq
|\Psi_{kl}\rangle = \int f(\beta_{30},\beta_{32}) \hat{P}_{(N,
Z)}|\Phi(\beta_{30}, \beta_{32})_{kl}\rangle d \beta_{30}d \beta_{32}
\eeq

%\emph{Changes}

We have restricted this study by imposing the quadrupole moment to
be fixed. A full calculation would require to consider the
octupole and quadrupole modes simultaneously, which would be a
huge computational task, well beyond the scope of the present
study. Our aim is indeed only to determine the most favourable
scenario of coupling between axial octupole mode and
the non axial $Q_{32}$ mode generating the tetrahedral deformation.

\begin{table}[H]
\begin{center}
\begin{tabular}{cccccccc}
\hline \hline
SLy4  &$E_{exc}$ (MeV)& $E_{corr}$ (MeV) &  $\pi$ & $\pi_{xy}$ &  $\tilde{\beta}_{30}$ & $\tilde{\beta}_{32}$ \\
\hline \hline
$^{110}$Zr& 0    & 1.222 & $+1$ &  $+1$& 0.08 & 0.06   \\
         & 4.535 &  -    & $-1$ &  $-1$& 0.07 & 0.17  \\
         & 4.282 &  -    & $-1$ &  $+1$& 0.15 & 0.04   \\
         & 7.423 &  -    & $+1$ &  $-1$& 0.19 & 0.24   \\
\hline
$^{154}$Gd& 0     & 2.228 &$+1$ & $+1$ & 0.05 & 0.06  \\
          & 2.892 &  -    &$-1$ & $-1$ & 0.06 & 0.10   \\
          & 1.998 &  -    &$-1$ & $+1$ & 0.12 & 0.04   \\
          & 5.005 &  -    &$+1$ & $-1$ & 0.12 & 0.10   \\
\hline \hline
\end{tabular}
\end{center}
\caption{Excitation energies, correlation energies and dynamical deformations of the lowest four
states obtained in 2-dim GCM.  $\pi$ and $\pi_{xy}$ denote the
parity and $P_{xy}$ quantum numbers, respectively.}
\end{table}

We have first performed a calculation in the vicinity of the
deformed ground state of two nuclei, $^{110}$Zr and
$^{154}$Gd. The results are shown in Table II.

The correlation energy due to the octupole modes is defined by:
\begin{equation*}
    E_{corr} = E (N,Z,sph) - E_{++},
\end{equation*}
where $E (N,Z,sph)$ is the energy of the particle number projected
spherical configuration and $E_{++}$ is the lowest positive-parity
energy obtained in the GCM. The value of this correlation energy
is rather small for both nuclei, indicating a weak effect of
octupole correlations in the ground state.

The lowest octupole excitation corresponds in both cases to the
axial octupole mode. The non axial octupole excitation is only
slightly larger in energy for $^{110}$Zr but both modes being
above 4~MeV of excitation are very unlikely to survive to the
coupling to any other modes. The situation is slightly more
favorable in $^{154}$Gd, although in this case, the non axial
excitation is nearly 1~MeV above the axial octupole one. Note that
in the vicinity of a deformed ground state, one cannot identify a
non axial $Q_{32}$ mode with tetrahedral deformations, since the
tetrahedral symmetry is broken by large quadrupole deformations.

In view of the very unfavorable conditions obtained when the
quadrupole moment is large, we have continued this study by
looking to the octupole properties around the spherical
configuration. The GCM results are summarized in  Table III. We
have performed calculations with two sets of mean-field wave
functions corresponding to 16 and 25 positive octupole
deformations respectively, to check the accuracy of the results.
The difference between both sets of results shows that the
accuracy obtained with a 25 wave-function basis set is better than
100~keV. Note that the excellent agreement obtained by using two
different basis sets is also a test that our results are not
affected by the pathology that can appear when working an energy
density functional\cite{BDH08}.

%\emph{Changes ended}

The largest gain is obtained for $^{110}Zr$. It is of a similar
order of magnitude than the energy gain due to quadrupole
correlations in deformed nuclei~\cite{bbh06}. A full study of the
energy gain due to the coupling between different modes remain to
be done but the results of Ref~\cite{bbh06} seem to indicate that
these energy gains quickly saturate in models based on
self-consistent mean-field wave functions.

Dynamical deformations associated with the lowest GCM
solutions corresponding to quantum numbers $k$ and $l$
are defined by:
\beq
\tilde{\beta}_{3\mu}=\sum_{\beta_{3\mu}} \beta_{3\mu}
g_{kl}^{2}(\beta_{30},\beta_{32}),
\eeq
for $\mu=0,2$, where $g_{kl}$ is the collective wave function for
parity $k$ and for an eigenvalue $l$ associated with the operator
$\hat{P}_{xy}$ (see Ref.~\cite{gcm} for the relation between the
collective wave function $g$ and the GCM function $f$). For all
nuclei that we have studied the dynamical deformations
$\tilde{\beta}_{30}$ and $\tilde{\beta}_{32}$ of the lowest
positive parity GCM solutions are smaller than 0.1. The ground
state collective wave function is rather isotropic as a function
of $Q_{30}$ and $Q_{32}$. It shows a similar spreading as a
function of axial and tetrahedral octupole deformations.

\begin{table}[H]
\begin{center}
\begin{tabular}{ccccccccccc}
\hline
\hline
 SIII &$E_{exc}$ (MeV)& $E_{corr}$ (MeV) & $\Delta_{16/25}$ & $\pi$ & $\pi_{xy}$ & $\tilde{\beta}_{30}$ & $\tilde{\beta}_{32}$ & $2 E_{qp}^{p}$ (MeV) & $2 E_{qp}^{n}$  (MeV)  \\
\hline
$^{80}$Zr& 0     & 2.116 & 0.15 &$+1$ & $+1$ & 0.07 & 0.06 &  2.786 & 3.518 \\
         & 2.832 &  -    & 0.06 &$-1$ & $-1$ & 0.05 & 0.18  & & \\
         & 2.854 &  -    & 0.01 &$-1$ & $+1$ & 0.14 & 0.00   & & \\
         & 8.275 &  -    & 0.16 &$+1$ & $-1$ & 0.13 & 0.17  & &  \\
\hline
$^{98}$Zr& 0.0   & 1.184 & 0.02 &$+1$ & $+1$& 0.07 & 0.04 &  2.114 & 2.66\\
         & 2.128 &  -    & 0.12 &$-1$ & $-1$& 0.04 & 0.25  & & \\
         & 1.732 &  -    & 0.09 &$-1$ & $+1$& 0.19 & 0.02  & & \\
         & 6.628 &  -    & 0.12 &$+1$ & $-1$& 0.17 & 0.23  & & \\
\hline
\hline
SLy4  &$E_{exc}$ (MeV)& $E_{corr}$ (MeV) & $\Delta_{16/25}$ & $\pi$ & $\pi_{xy}$ & $\tilde{\beta}_{30}$ & $\tilde{\beta}_{32}$ & $2 E_{qp}^{p}$ (MeV) & $2 E_{qp}^{n}$  (MeV)\\
\hline
$^{98}$Zr& 0     & 2.660 & 0.07&$+1$ & $+1$ & 0.10 & 0.08 & 1.96 & 1.78\\
         & 2.393 &  -    & 0.06&$-1$ & $-1$ & 0.08 & 0.21  & & \\
         & 2.639 &  -    & 0.05&$-1$ & $+1$ & 0.18 & 0.06  & & \\
         & 6.127 &  -    & 0.02&$+1$ & $-1$ & 0.17 & 0.17  & & \\
\hline
$^{110}$Zr& 0    & 3.303 & 0.01&$+1$ &  $+1$& 0.09 & 0.10  & 1.612 & 2.72 \\
         & 1.764 &  -    & 0.01&$-1$ &  $-1$& 0.06 & 0.22  & & \\
         & 2.188 &  -    & 0.01&$-1$ &  $+1$& 0.17 & 0.06  & & \\
         & 4.936 &  -    & 0.01&$+1$ &  $-1$& 0.16 & 0.20  & & \\
\hline
$^{152}$Gd& 0     & 2.791 & 0.00&$+1$ & $+1$ & 0.05 & 0.06  & 2.884 & 2.78 \\
          & 2.018 &  -    & 0.00&$-1$ & $-1$ & 0.04 & 0.13  & & \\
          & 2.233 &  -    & 0.00&$-1$ & $+1$ & 0.11 & 0.05  & & \\
          & 4.922 &  -    & 0.01&$+1$ & $-1$ & 0.12 & 0.12  & & \\
\hline
$^{154}$Gd& 0     & 3.054 & 0.00&$+1$ & $+1$& 0.06 & 0.07  & 2.566 & 3.0 \\
          & 1.507 &  -    & 0.01&$-1$ & $-1$& 0.05 & 0.14  & & \\
          & 1.857 &  -    & 0.01&$-1$ & $+1$& 0.12 & 0.05  & & \\
          & 4.134 &  -    & 0.00&$+1$ & $-1$& 0.11 & 0.02  & & \\
\hline
$^{156}$Gd& 0     & 3.085 & 0.10&$+1$ & $+1$& 0.06 & 0.08  & 2.008 & 2.742 \\
          & 1.072 &  -    & 0.00&$-1$ & $-1$& 0.05 & 0.15  & & \\
          & 1.507 &  -    & 0.06&$-1$ & $+1$& 0.12 & 0.05  & & \\
          & 3.329 &  -    & 0.02&$+1$ & $-1$& 0.11 & 0.13  & & \\
\hline
$^{160}$Yb& 0     & 3.085 & 0.00&$+1$ & $+1$ & 0.06 & 0.06  & 3.438 & 3.06 \\
          & 1.629 &  -    & 0.00&$-1$ & $-1$ & 0.05 & 0.13  & & \\
          & 1.858 &  -    & 0.02&$-1$ & $+1$ & 0.11 & 0.05  & & \\
          & 3.893 &  -    & 0.02&$+1$ & $-1$ & 0.10 & 0.12  & & \\
\hline
\hline
\end{tabular}
\end{center}
\caption{Excitation energies, correlation energies and dynamical deformations of the lowest four
states obtained in 2-dim GCM. $\Delta_{16/25}$ denotes the
difference in energies between calculations performed with 16 and
25 mean-field states. $\pi$ and $\pi_{xy}$ denote the parity and
$P_{xy}$ quantum numbers, respectively. In the last two columns
the two-quasiparticle excitation energies (neutron and proton) of
the spherical configuration are given.}
\end{table}

The first negative parity state has an excitation energy comprised
between $1.0$ and $2.3$ MeV, the tetrahedral mode being
systematically the lowest one. The largest differences
$E_{+-}-E_{--}$ between both octupole modes occur for $^{110}$Zr
where it is around  0.8~MeV and for $^{156}$Gd where it is around
0.5~MeV.

The ratio between the B(E3) values obtained for both modes are
given in Table IV. Better than the absolute values of these
quantities which are not well defined in an angular momentum
unprojected model, these ratios are good indicators whether these
states have a specific signature in their deexcitation spectrum.
In the first column are given the ratios corresponding to the
transitions  from the tetrahedral and the axial excited states to
the ground state. The second column corresponds to the ratios of
the transitions between the mixed octupole states to the
tetrahedral and the axial excitations. This ratio oscillates in
all cases around $1$ which suggests that the spectrum of GCM
excitations resembles to a large extent a harmonic spectrum. The
only noticeably deviation occurs in the case of $^{110}$Zr where
the transition from the tetrahedral  state to the lowest GCM state
is decreased by about 30\% as compared to the deexcitation of the
axial octupole vibration.

\begin{table}[H]
\begin{center}
\begin{tabular}{ccccccccc}
\hline
\hline
 SLy4 &a & b  \\
\hline
\hline
$^{98}$Zr & 0.78   & 1.05   \\

$^{110}$Zr& 0.67  & 0.72  \\

$^{152}$Gd& 0.83   & 0.90  \\

$^{154}$Gd& 0.97    & 0.91 \\

$^{156}$Gd& 1.15    & 0.89 \\

$^{160}$Yb& 1.04    & 1.12 \\
\hline
\hline
\end{tabular}
\end{center}
\caption{Ratios of the B(E3) values obtained for the transitions between the four lowest GCM states. In the column denoted by $a$,
the ratio is taken between $|\Phi_{--} \rangle \rightarrow
|\Phi_{++} \rangle $, and $|\Phi_{-+} \rangle \rightarrow
|\Phi_{++} \rangle $, and in $b$ for: $|\Phi_{+-} \rangle
\rightarrow |\Phi_{--} \rangle $ and  $|\Phi_{+-} \rangle
\rightarrow |\Phi_{-+} \rangle $.}
\end{table}

\section{CONCLUSIONS}

We have investigated the possible existence of stable tetrahedral
configurations in nuclei in which they were predicted on the
basis of non self-consistent models. Our calculations have been
based on several parametrization of the Skyrme interaction, with only
marginal differences between the results. The coupling between the axial
and tetrahedral octupole modes has been studied with the GCM, in the
absence of quadrupole deformations. Our results do not
support the prediction that tetrahedral deformations should have a definite signature:

\begin{itemize}
\item The susceptibility of the spherical
configuration towards tetrahedral deformations is rather weak and
pairing effects wash out the shell effects. Moreover the
tetrahedral minimum is accompanied by an axial octupole minimum of
similar depth.

\item The correlation energy associated with shape fluctuations
and parity restoration lowers  substantially the mean-field
energy. However the dynamic octupole deformations in the ground
state state is rather small.
Moreover axial and non axial octupole deformations are strongly
coupled.

\item The excitation energies of states associated with tetrahedral shapes
are systematically lower than those corresponding to the axial octupole mode.
However the B(E3) ratios do not distinguish between these modes.
\end{itemize}

The prospects for the experimental detection of the tetrahedral
configurations at spin zero are thus rather poor. It seems that
the increased shell effects due to the tetrahedral mode do not
provide a sufficient condition for the existence of a stable
tetrahedral deformation. At spin zero, stable tetrahedral
configurations seem unlikely. Their trace may be manifested in
nuclear vibrations in negative parity bands but the B(E3) values
indicate that there is no way to distinguish the tetrahedral modes
from the axial octupole modes by looking to the decay
probabilities. It should be noted that our study does not rule out
the possibility of the existence of rotating tetrahedral
configurations. Several of the predicted tetrahedral nuclei are
however strongly deformed in their ground state and a mixing of
octupole and quadrupole deformations would make still more
problematic the extraction of a tetrahedral signature.

\begin{acknowledgments}
Discussions with M. Bender, J. Dobaczewski, P. Olbratowski, W. Satu{\l}a and J. Skalski
are gratefully acknowledged. 
This work has been supported in part by the Polish Ministry of
Science under contract No. N N202 328234, the
Foundation for Polish Science (FNP), the PAI P6-23 of the
Belgian Office for Scientific Policy and the US Department
of Energy Grant No. DE-FC02-07ER41457.

Numerical calculations were performed at the Interdisciplinary Centre
for Mathematical and Computational Modelling (ICM) at Warsaw University.
\end{acknowledgments}


\begin{thebibliography}{99}
%1
\bibitem{bn}  P. Butler and W. Nazarewicz,
              Rev. Mod. Phys. {\bf 68} 349 (1996).
\bibitem{ska92} J. Skalski, Phys. Lett {\bf B274} 1 (1992)
\bibitem{bfh87} P. Bonche, P.-H.H.  Heenen, Flocard and
                D. Vautherin, Phys. Lett. {\bf B175}  387 (1986)
\bibitem{er89}  L. Egido and L. Robledo Nucl. Phys. {\bf A494} 85 (1989)
\bibitem{eb03}  J. Engel, M. Bender, J. Dobaczewski, J.H. Jesus and P. Olbratowski,
                Phys. Rev. {\bf C68} 025501 (2003)
\bibitem{mbw95} J. Meyer, P. Bonche, M. Weiss, H. Flocard and P.-H. Heenen,
                Nucl. Phys. {\bf A588} 597 (1995)
\bibitem{ld}  X. Li and J. Dudek,
              Phys. Rev. {\bf C49} R1250 (1994).
\bibitem{dgs} J. Dudek, A. G\'o\'zd\'z, N. Schunck, M. Mi\'skiewicz,
              Phys. Rev. Lett. {\bf 88} 252502 (2002);
              N. Schunck, J. Dudek, A. G\'o\'zd\'z and P.H. Regan,
              Phys. Rev. {\bf C69} 061305(R) (2004).
\bibitem{dgs2} J. Dudek, A. G\'o\'zd\'z and N. Schunck,
               Acta Phys.Polon. {\bf B34} 2491 (2003); N. Schunck, J. Dudek,
               Int.J.Mod.Phys. {\bf E13} 213 (2004).
\bibitem{sod}  N. Schunck, P. Olbratowski, J. Dudek, J. Dobaczewski,
               Int.J.Mod.Phys. {\bf E15} 490 (2006).
\bibitem{dcd} J. Dudek, D. Curien, N. Dubray, J. Dobaczewski, V. Pagnon, P. Olbratowski and N. Schunck,
              Phys. Rev. Lett. {\bf 97}, 072501 (2006).
\bibitem{ddd} J. Dudek, J. Dobaczewski, N. Dubray, A. Gozdz, V. Pangon, N. Schunck,
               Int.J.Mod.Phys. {\bf E16} 516 (2007).
\bibitem{zber1} K. Zberecki, P. Magierski, P.-H. Heenen, N. Schunck, Phys. Rev. {\bf C74} 051302(R) (2006).
\bibitem{zber2} K. Zberecki, P. Magierski, P.-H. Heenen, N. Schunck, Int. J. Mod. Phys. {\bf E16}  533 (2007) 533.
\bibitem{gcm} M. Bender, P.-H. Heenen, and P.-G. Reinhard,
              Rev. Mod. Phys. {\bf 75} 121 (2003).
\bibitem{lcc87} C. J. Lister, M. Campbell, A. A. Chishti, W. Gelletly, L.
Goettig, R. Moscrop, B. J. Varley, A. N. James, T. Morrison, H. G.
Price, J. Simpson, K. Connel, and O. Skeppstedt, Phys. Rev. Lett. {\bf 59},
1270 (1987).
\bibitem{shb93a} J. Skalski, P.-H. Heenen, P. Bonche, H. Flocard and J. Meyer,
                 Nucl. Phys. {\bf A551} 109 (1993).
\bibitem{dmn} J. Dobaczewski, P. Magierski, W. Nazarewicz, W. Satu{\l}a and
              Z. Szyma\'nski, Phys. Rev. {\bf C63} 024308 (2001).
\bibitem{BDH08} D. Lacroix, T. Duguet and M. Bender, arXiv:0809.2041v1
\bibitem{bbh06} M. Bender, G. F. Bertsch, and P.-H. Heenen,
                Phys. Rev. C \textbf{73}, 034322 (2006).



\end{thebibliography}
\end{document}